\documentclass[english,aps,preprint,superscriptaddress]{revtex4-1}
\usepackage[T1]{fontenc}
\usepackage[latin9]{inputenc}
\usepackage{units}
\usepackage{amsmath}
\usepackage{graphicx}
\usepackage{color}
\usepackage{babel}
\usepackage{textcomp}
\usepackage{mathrsfs}
\usepackage{dsfont}
\usepackage{amssymb}
\usepackage[unicode=true,pdfusetitle,
 bookmarks=true,bookmarksnumbered=false,bookmarksopen=false,
 breaklinks=false,pdfborder={0 0 0},pdfborderstyle={},backref=false,colorlinks=true]
 {hyperref}

\begin{document}
\title{Quark and gluon entanglement in the proton based on a light-front Hamiltonian}

\author{Chen Qian}
\email{qianchen@baqis.ac.cn}
\affiliation{Beijing Academy of Quantum Information Sciences, Beijing 100193,
China}

\author{Siqi Xu}
\email{xsq234@163.com}
\affiliation{Institute of Modern Physics, Chinese Academy of Sciences,
Lanzhou 730000, China}
\affiliation{School of Nuclear Science and Technology, University of Chinese
Academy of Sciences, Beijing 100049, China}
\affiliation{Department of Physics and Astronomy, Iowa State University, Ames, Iowa 50011, USA}

\author{Yang-Guang Yang}
\email{yyg@impcas.ac.cn}
\affiliation{Institute of Modern Physics, Chinese Academy of Sciences,
Lanzhou 730000, China}

\author{Xingbo Zhao}
\email{xbzhao@impcas.ac.cn}
\affiliation{Institute of Modern Physics, Chinese Academy of Sciences,
Lanzhou 730000, China}
\affiliation{School of Nuclear Science and Technology, University of Chinese
Academy of Sciences, Beijing 100049, China}

\begin{abstract}
Given that the wave function of a proton can be derived relativistically and nonperturbatively from a light-front quantized Hamiltonian, investigating the quantum correlation between quarks and gluons offers a novel perspective on the internal structure of partons within a proton. In this work, we address this topic by computing the spin and longitudinal momentum entanglement of each parton inside the proton. The utilized wave functions are generated using Basis Light-front Quantization (BLFQ), incorporating both the valence quarks and one dynamical gluon Fock sectors, $\left|qqq\right\rangle$ and $\left|qqq\right\rangle +\left|qqqg\right\rangle$. Our calculations indicate that the dynamical gluon significantly enhances entanglement among the proton's partons. Additionally, we examine the spin entanglement of quarks and gluons at fixed values of longitudinal momentum fraction, revealing that the presence of a gluon may amplify the informational exchanges between quarks. Finally, these findings suggest the potential for experimental verification of the entanglement between partons by measuring parton helicity distributions in the proton.
\end{abstract}
\maketitle

\section{Introduction}

Quantum nonlocality, a remarkable phenomenon in quantum mechanics, is typically characterized by the violation of Bell inequalities and other entanglement measures~\cite{Brunner:2014rmp, Horodecki:2009rmp}. Previously, atomic and quantum optic system experiments have successfully explored these phenomena~\cite{Yin:2017ips, BIGBellTest:2018ebd}. In recent years, the properties of quantum nonlocality in elementary particle physics have attracted significant attention, offering a fresh perspective on studying fundamental interactions~\cite{Afik:2022kwm, ATLAS:2023fsd, Wu:2024mtj, Wu:2024asu}. Among these developments, the study of quantum correlations between colored partons within a hadron has emerged as a promising topic.
Researches involving deep inelastic scattering has demonstrated parton entanglement~\cite{Kharzeev:2017qzs, Tu:2019ouv, Gotsman:2020bjc, Kharzeev:2021yyf, Zhang:2021hra}, which have been observed through experimental data as well~\cite{Tu:2019ouv, Kharzeev:2021yyf}. In these studies, entanglement entropy is commonly employed to quantify the entanglement between partons, which segments a proton's pure state
into two subregions and can be expressed as a function of the light cone momentum fractions $x$. This topic has sparked widespread interest in studying quantum correlation properties in quantum chromodynamics (QCD) and strong interactions~\cite{Ehlers:2022oal, Hentschinski:2022rsa, Wang:2014lua, Beane:2018oxh, Liu:2022grf, Miller:2023ujx}. Recent investigations have utilized entanglement entropy in perturbative methods to explore the spin and angular correlations of partons inside proton wave functions~\cite{Dumitru:2022tud, Dumitru:2023qee, Ehlers:2022oal}.

The framework of Basis Light-front Quantization (BLFQ) provides a relativistic nonperturbative approach for addressing many-body problems in quantum field theory~\cite{Vary:2009gt}. Based on the light-front Hamiltonian formalism, BLFQ enables us to extract comprehensive information about hadron systems on the amplitude level. Thus, the wave functions from BLFQ serve as an important tool for investigating the quantum nonlocality and entanglement in the internal structure of hadrons.

In the previous work~\cite{Mondal:2019jdg, Xu:2021wwj, Xu:2023nqv}, the authors expand the proton state to the valence Fock sector ($\left|qqq\right\rangle$) and one-dynamical gluon Fock sector ($\left|qqq\right\rangle +\left|qqqg\right\rangle$) and solve the eigenvalue equation from the light-front effective Hamiltonian, which includes the quark-gluon interactions from QCD and a three-dimensional confining potential. The resulting wave functions of BLFQ provide a reasonable description of the proton structures~\cite{Mondal:2019jdg, Xu:2021wwj, Xu:2023nqv, Hu:2020arv, Liu:2022fvl, Hu:2022ctr, Kaur:2023lun, Lin:2023ezw}. In this work, we study proton wave function entanglement structures produced by Basis Light-front Quantization (BLFQ)~\cite{Mondal:2019jdg, Xu:2021wwj, Xu:2023nqv}. Utilizing the light-front wave functions, we first examine the entanglement entropy associated with the spin and longitudinal momentum of quarks and gluons. Subsequently, we examine the quantum nonlocality of quark spin through the violation of the Bell-CH inequality. We observe the dependence of the entanglement entropy on the expansion of Fock space, which suggests that entanglement entropy may serve as a valuable measure for characterizing the structures of these bound states. Meanwhile, our analysis reveals only a marginal violation of the Bell-CH inequality in the state $\left|qqq\right\rangle$, indicating that the nonlocal correlations among the three quarks might be weak. Furthermore, to make the connection with the experimental observables, we calculate the entanglement entropy associated with spin across various configurations of longitudinal momentum fractions $x$ for both quarks and gluons. These results provide avenues for investigating the entanglement spectrum of the proton through scattering experiments, such as those performed at the hadron and electron-ion colliders~\cite{CMS:2010qvf}.

This paper is organized as follows. We briefly review the concept of the Basis Light-front Quantization (BLFQ), entanglement entropy, and Bell quantum nonlocality in Sec.~\ref{sec:Preliminaries}. Subsequently, we show the numerical results of the entanglement entropy and violation of three-body Bell-CH inequality from the BLFQ wave function in Sec.~\ref{sec:Nonlocality-and-entanglement}. After that, we discuss the entanglement entropy of quark and gluon spin in terms of longitudinal momentum fractions in Sec.~\ref{sec:Spin-entanglement}. Finally, we summarize the main results and give an outlook in Sec.~\ref{sec:Conclusions-and-outlook}.

\section{Methodology\label{sec:Preliminaries}}

This section begins with a brief overview of the proton wave function derived from the Basis Light-front Quantization (BLFQ) framework~\cite{Vary:2009gt, Xu:2023nqv}. After that, we introduce concepts of entanglement and Bell nonlocality, which we will apply to the BLFQ proton wave functions in the following sections.

\subsection{Basis Light-front Quantization for proton}\label{subsec:BLFQ}

Basis Light-front Quantization is a nonperturbative framework for solving the relativistic bound state in quantum field theory. By transforming the Minkowski coordinates $\left(x^{0},x^{1},x^{2},x^{3}\right)$ into light-front coordinates $\left(x^{+},x^{1},x^{2},x^{-}\right)$ $\left(x^{\pm}=x^{0}\pm x^{3}\right)$, and quantizing the field theory on the light front (the hypersurface with equal light-front time $x^+$). We can formulate the light-front Hamiltonian of QCD with the momentum $P^-_{QCD}$ ($P^\pm=P^0\pm P^3$). The information of bound states is encoded by the light-front wave functions (LFWFs) which are generated by diagonalizing the light-front Hamiltonian:
\begin{equation}
    P^-_{QCD}P^+\left|\Psi\right\rangle =M^{2}_\Psi\left|\Psi\right\rangle ,\ \label{eq:lfeq}
\end{equation}
where $M^2_\Psi$ is the eigenvalue corresponding to the mass squared, and the associated eigenvector $\left|\Psi\right\rangle$ encodes the structural information of the proton state. 
At fixed light-front (LF) time $x^{+}= 0$, the state of a baryon can be expressed in Fock space as
\begin{equation}
\left|\Psi\right\rangle =\varphi_{(qqq)}\left|qqq\right\rangle +\varphi_{(qqqg)}\left|qqqg\right\rangle + \varphi_{(qqqq\bar{q})}\left|qqqq\bar{q}\right\rangle \cdots,\ \label{eq:baryon-wave}
\end{equation}
where $\varphi_{(\cdots)}$ is the LFWFs related to the different Fock sectors $\left|qqq\cdots\right\rangle$. 
To represent the BLFQ wave functions, we use a two-dimensional harmonic oscillator (2D-HO) basis in the transverse directions, specified by the radial quantum number $n$ and the orbital angular momentum quantum number $m$. Additionally, we adopt plane-wave states confined within a one-dimensional box of length $2L$, with periodic (antiperiodic) boundary conditions for the longitudinal degrees of freedom for quarks (gluons), characterized by the longitudinal quantum number $k$. Including the light-cone helicity $\lambda$, every single parton in a given Fock sector is described by the four quantum numbers $\alpha = \{k, n, m, \lambda\}$. Each Fock sector also includes an additional quantum number \( c \), which represents the color degree of freedom. Furthermore, we introduce two truncation parameters: $\sum_{i} \left( 2n_{i} + |m_{i}| + 1 \right) \leq N_{\text{max}}$ and $\sum_{i} k_{i} = K$. In this work, we study the quantum correlation in light-front wave functions which are produced in BLFQ from two model Hamiltonians in the Fock space truncated to the $\left|qqq\right\rangle$~\cite{Mondal:2019jdg, Xu:2021wwj} and $\left|qqq\right\rangle+\left|qqqg\right\rangle$~\cite{Xu:2023nqv} Fock sectors respectively.

\subsubsection{Truncation to the $\left|qqq\right\rangle$ Fock sector}

In the framework of BLFQ, the proton can be interpreted as a wave function with different numbers of partons by the Fock-space expansion. The leading Fock sector only includes the valence quarks ($\left|uud\right\rangle$). In the truncated Fock space with only the leading Fock sector, we work with an effective light-front Hamiltonian with three terms~\cite{Mondal:2019jdg, Xu:2021wwj}:
\begin{eqnarray}
    P_{QCD}^- & = & \int d^{2}x^{\perp}dx^{-} \frac{1}{2}\bar{\Phi}\gamma^{+}\frac{m_{0}^{2}+\left(i\partial^{\perp}\right)^{2}}{i\partial^{+}}\Phi + P^-_C + P^-_{OGE}, \label{eq:effective-hamiltonian}
\end{eqnarray}
where the $\Phi$ is the quark field. The first term in the above equation is the kinetic term. $P^-_C$ represents the confinement potential and $P^-_{OGE}$ provides the effective strong interaction between the valence quarks. For the three valence quarks, the confinement potential can be written as:
\begin{eqnarray}
    P_{C}^{-}P^{+} & = & \frac{\kappa^{4}}{2}\sum_{i\neq j}^{3}\left\{ \vec{r}_{ij\perp}^{2}-\frac{\partial_{x_{i}}\left(x_{i}x_{j}\partial_{x_{j}}\right)}{\left(m_{i}+m_{j}\right)^{2}}\right\} ,\ \label{eq:confining-interaction}
\end{eqnarray}
where $m_i$ is the physical quark mass and $\kappa$ is the strength of the confinement. Here we adopt the light-front soft-wall model~\cite{deTeramond:2008ht} for the confining potential in the transverse plane and augment it with a longitudinal confining potential equivalent to the one-dimensional harmonic oscillator potential in the non-relativistic limit~\cite{Li:2015zda}. The $\vec{r}_{ij\perp}=\sqrt{x_{i}x_{j}}\left(\vec{r}_{i\bot}-\vec{r}_{j\bot}\right)$ represents the intrinsic coordinates between the valence quarks. It is worth mentioning that the longitudinal confinement potential has been implemented in the heavy quarkonium system in relativistic regimes~\cite{Li:2015zda,Li:2017mlw}.

The gluon is a gauge boson that acts as a mediator of the strong interaction. In the basis truncated to the leading Fock sector, we employed the one-gluon exchange interaction given by:
\begin{align}
    P^-_{{OGE}}P^+ = \Sigma_{i\neq j}\frac{4\pi C_F \alpha_s}{Q^2_{ij}} \bar{u}_{s^{\prime}_i}(p^{\prime}_i)\gamma^{\mu}u_{s_i}(p_i)\bar{u}_{s^{\prime}_j}(p^{\prime}_j)\gamma_{\mu}u_{s_j}(p_j),
\end{align}
with fixed coupling constant $\alpha_s$. Here, $Q^2_{ij}=-(1/2)(p^{\prime}_i-p_i)^2-(1/2)(p^{\prime}_j-p_j)^2$ is the average of 4-momentum square carried by the exchanged gluon. The color factor $C_F=-2/3$, which implies that the OGE is an attractive potential, can be derived from the $\mathrm{SU\left(3\right)}$ group theory. Implementing the OGE interaction, we get the dynamical spin structure in the LFWFs, which is crucial in computing the spin-dependent observables.

In the $\left|qqq\right\rangle$ case, we select the truncation parameters $N_{\rm{max}}=10$ and $K_{\rm{max}}=16.5$. The model parameters are summarized in Table~\ref{tab:parameter}, where $m_{\rm{q/k}}$ and $m_{\rm{q/g}}$ denote the mass of quarks in kinetic terms and interaction terms respectively. These parameters are determined by fitting the proton mass and the flavor Dirac form factors~\cite{deTeramond:2008ht}.
\begin{table}[htp]	
    \centering
    \caption{List of the model parameters for the basis truncation's $N_{\rm{max}}=10$ and $K_{\rm{max}}=16.5$.}\label{tab:parameter}
    \begin{tabular}{cccccc}
    \hline\hline
    $m_{\rm{q/k}}$ 	   ~&~   $m_{\rm{q/g}}$    ~&~ $\kappa$      	 ~&~ $\alpha_s$    \\
    \hline
    $0.3$ GeV ~&~	 $0.2$	GeV    ~&~ $0.34$ GeV ~&~ $1.1\pm 0.1$ \\
    \hline\hline
    \end{tabular}
    \end{table} 

\subsubsection{Truncation to the $\left|qqq\right\rangle+\left|qqqg\right\rangle$ Fock sector}

In the basis space of $\left|qqq\right\rangle+\left|qqqg\right\rangle$, the proton can be expressed in terms of the valence quarks $\left|uud\right\rangle$ Fock sector and three quarks with one dynamical gluon $\left|uudg\right\rangle$.
In this truncated Fock space, we adopt the model of light-front Hamiltonian $P^{-}=P_{QCD}^{-}+P_{C}^{-}$, where $P_{QCD}^{-}$ denotes the LF QCD Hamiltonian that includes interactions relevant to the two leading Fock sectors $\left|uud\right\rangle$ and $\left|uudg\right\rangle$, and $P_{C}^{-}$ refers to a model for the confining interaction~\cite{Xu:2023nqv}. In LF gauge ($A^+=0$), the form of $P_{QCD}^{-}$ can be written as 
\begin{eqnarray}
P_{QCD}^{-} & = & \int d^{2}x^{\perp}dx^{-}\left\{ \frac{1}{2}\bar{\Phi}\gamma^{+}\frac{m_{0}^{2}+\left(i\partial^{\perp}\right)^{2}}{i\partial^{+}}\Phi\right.\nonumber \\
 &  & -\frac{1}{2}A_{a}^{i}\left[m_{g}^{2}+\left(i\partial^{\perp}\right)^{2}\right]A_{a}^{i}+g_{S}\bar{\Phi}\gamma_{\mu}T^{a}A_{a}^{\mu}\Phi\nonumber \\
 &  & \left.+\frac{1}{2}g_{S}^{2}\bar{\Phi}\gamma^{+}T^{a}\Phi\frac{1}{\left(i\partial^{+}\right)^{2}}\bar{\Phi}\gamma^{+}T^{a}\Phi\right\} ,\ \label{eq:P-QCD}
\end{eqnarray}
where $\Phi$ and $A^{\mu}$ represent the quark and gluon fields, $x^{-}$ and $x^{\perp}$ are the longitudinal and transverse position
coordinates, respectively. $T^{a}$ denotes the generator of $\mathrm{SU\left(3\right)}$ gauge group in color space, and $\gamma_{\mu}$ are the Dirac matrices. The first two terms of Eq.~$\left(\ref{eq:P-QCD}\right)$ represent the kinetic energies associated with the quark and gluon, possessing bare masses $m_{0}$ and $m_{g}$, respectively. Following the Fock-sector dependent renormalization theory, we produce the mass counter term ($\delta m$) and define the bare mass $m_{0}=m_{q}+\delta m_{q}$ in the leading Fock sector, where $m_{q}$ denotes the physical quark mass. 
Complementing the confining potential in the leading Fock sectors shown below, we adopt an effective mass of gluon $m_{g}$ to model the phenomenological confinement in the $\left|qqqg\right\rangle$ Fock sector. The remaining terms in Eq.~$\left(\ref{eq:P-QCD}\right)$ are the vertex and instantaneous interactions with the strong coupling constant $g_{S}$. 

In the leading Fock sector we adopt the same transverse and longitudinal confining potential as the in the previous model, cf. Eq.~$\left(\ref{eq:effective-hamiltonian}\right)$, as follows~\cite{Li:2015zda, Lan:2021wok}
\begin{eqnarray}
P_{C}^{-}P^{+} & = & \frac{\kappa^{4}}{2}\sum_{i\neq j}^{3}\left\{ \vec{r}_{ij\perp}^{2}-\frac{\partial_{x_i}\left(x_{i}x_{j}\partial_{x_{j}}\right)}{\left(m_{i}+m_{j}\right)^{2}}\right\} \; \label{eq:confining-potential},
\end{eqnarray}
where $\kappa$ is the strength of confinement. $\vec{r}_{ij\perp}=\sqrt{x_{i}x_{j}}\left(\vec{r}_{i\bot}-\vec{r}_{j\bot}\right)$ represents the separation between the struck parton and the spectator.

In the explicit truncation parameters $\left\{ N_{max}, K\right\} =\left\{ 9,16.5\right\}$, we choose HO scale parameter $b=0.7\mathrm{\;GeV}$ and UV cutoff for the instantaneous interaction $b_\mathrm{inst}=3\mathrm{\;GeV}$, and set our model parameters $\left\{ m_{u},m_{d},m_{g},\kappa,m_{f},\tilde{g}_{S}\right\} =\left\{ 0.31~\mathrm{GeV},0.25~\mathrm{GeV},0.5~\mathrm{GeV},0.54,1.8~\mathrm{GeV},2.4\right\}$ by fitting the proton mass and electromagnetic form factors. Here, in order to model the nonperturbative effects from the higher Fock sectors, we treat the quark mass in the vertex interaction, $m_f$, as an independent phenomenological parameter~\cite{Xu:2023nqv, Glazek:1992aq, Burkardt:1998dd}.

\subsection{Entanglement entropy}\label{subsec:EE}

After introducing the proton wave function obtained from Basis Light-front Quantization, we review quantities to measure quantum correlations between partons inside protons. Here, we begin with entanglement entropy, which is regarded as a good measure of entanglement between two parties of one pure state. For a pure state in $N$-partite system $\left|\psi\right\rangle$, its $N\times N$ density matrix can be written as $\rho_{\mathrm{AB}}=\left|\psi\right\rangle \left\langle \psi\right|$. We divide the $N$-partite system into two parts, one is the part of our interest $A$, and the complementary part is $B$. The reduced density matrix can be derived by applying a partial trace
\begin{equation}
\rho_{\mathrm{A}}=\mathrm{Tr_{B}}\rho_{\mathrm{AB}},\:\rho_{\mathrm{B}}=\mathrm{Tr_{A}}\rho_{\mathrm{AB}},\;\label{eq:trace}
\end{equation}
where $\rho_{\mathrm{A}}$ and $\rho_{\mathrm{B}}$ are called reduced density matrices of subsystems $\mathrm{A}$ and $\mathrm{B}$, respectively. The entanglement between the two parts can be represented by Von-Neumann entanglement entropy, which reads
\begin{equation}
S=\mathrm{Tr}\left(\rho_{\mathrm{A}}\mathrm{\log_{2}\rho_{A}}\right)=\mathrm{Tr}\left(\rho_{\mathrm{B}}\mathrm{\log_{2}\rho_{B}}\right).\;\label{eq:ent-entropy}
\end{equation}
Due to the positivity of density matrix, Eq.~$\left(\ref{eq:ent-entropy}\right)$ is equivalent to the Shannon entropy
\begin{equation}
S=-\sum_{i=1}^{N_{\mathrm{A}}}p_{\mathrm{A}_{i}}\log_{2}p_{\mathrm{A}_{i}}=-\sum_{j=1}^{N_{\mathrm{B}}}p_{\mathrm{B}_{j}}\log_{2}p_{\mathrm{B}_{j}},\;\label{eq:Shannon-entropy}
\end{equation}
where $N_{\mathrm{A}}$ and $N_{\mathrm{B}}$ are dimensions of the subsystems $\mathrm{A}$ and $\mathrm{B}$, together with $\left\{ p_{\mathrm{A}_{i}}\right\} $ and $\left\{ p_{\mathrm{B}_{j}}\right\}$ being eigenvalues of $\rho_{\mathrm{A}}$ and $\rho_{\mathrm{B}}$. Entanglement entropy is a good measure of the entanglement between two complementary parts of any pure state since it is a monotonous function with $S=0$ when $\mathrm{A}$ and $\mathrm{B}$ are separable, and $S=\mathrm{min}\{N_{\mathrm{A}}, N_{\mathrm{B}}\}$ when $\mathrm{A}$ and $\mathrm{B}$ are maximally entangled.

\subsection{Quantum nonlocality and Bell-CH inequalities}\label{subsec:Bell}

Bell nonlocality is a stronger nonlocal correlation than entanglement in quantum information theory. It was introduced by John Bell in 1964 to refute the local hidden variable theory (LHVT)~\cite{Bell:1964kc}, typically evidenced through the violation of Bell-type inequalities. The premise of the Bell inequality is that if any local hidden variables can adequately explain quantum correlations, the inequality will hold. Conversely, violation of this inequality in systems exhibiting quantum correlations establishes the nonlocal nature of such correlations. To facilitate empirical validation within the field of quantum optics, Clauser and Horne advanced the theoretical groundwork by developing the Bell-CH inequalities, which are specifically designed for experimental testability~\cite{Clauser:1974tg}. These inequalities are defined in terms of probability distributions, making them directly applicable in experiments. 

Unlike entanglement entropy revealing the correlation between one specific subsystem and its complementary part, Bell-CH inequalities characterize a general nonlocality among the whole system. Here, we take the three valence quarks as one system and apply three-body Bell-CH inequalities to them. Specifically, Bell-CH inequalities are not unique, they can be classified by different types~\cite{Pitowsky2001, Rosset2014, Qian:2020ini, Qian:2021bmm}. Based on previous experience~\cite{Pitowsky2001}, in this work, we specifically consider a three-body Bell-CH inequality, which has demonstrated a large violation in nonlocally correlated quantum systems compared to its counterparts, thereby suggesting its efficacy in capturing nonlocal correlations~\cite{Pitowsky2001}. This Bell-CH inequality has the form of
\begin{eqnarray}
-3P\left(\mathbf{\mathbf{\boldsymbol{a}}_{\mathrm{1}}}\right)-2P\left(\mathbf{\mathbf{\boldsymbol{b}}_{\mathrm{1}}}\right)-P\left(\mathbf{\mathbf{\boldsymbol{c}}_{\mathrm{1}}}\right)+2P\left(\mathbf{\boldsymbol{a}}_{1},\mathbf{\boldsymbol{b}_{\mathrm{1}}}\right)+3P\left(\mathbf{\boldsymbol{a}_{\mathrm{1}}},\mathbf{\boldsymbol{b}_{\mathrm{2}}}\right)+2P\left(\mathbf{\boldsymbol{a}_{\mathrm{2}}},\mathbf{\boldsymbol{b}_{\mathrm{1}}}\right)\nonumber \\
-2P\left(\mathbf{\boldsymbol{a}_{\mathrm{2}}},\mathbf{\boldsymbol{b}_{\mathrm{2}}}\right)+P\left(\mathbf{\boldsymbol{a}_{\mathrm{1}}},\mathbf{\boldsymbol{c}_{\mathrm{1}}}\right)+P\left(\mathbf{\boldsymbol{a}_{\mathrm{2}}},\mathbf{\boldsymbol{c}_{\mathrm{1}}}\right)+P\left(\mathbf{\boldsymbol{b}_{\mathrm{1}}},\mathbf{\boldsymbol{c}_{\mathrm{1}}}\right)+P\left(\mathbf{\mathbf{\boldsymbol{a}}_{\mathrm{1}}\mathrm{,}\boldsymbol{b}_{\mathrm{1}}},\mathbf{\boldsymbol{c}_{\mathrm{1}}}\right)-2P\left(\mathbf{\mathbf{\boldsymbol{a}}_{\mathrm{2}}\mathrm{,}\boldsymbol{b}_{\mathrm{1}}},\mathbf{\boldsymbol{c}_{\mathrm{1}}}\right)\nonumber \\
+P\left(\mathbf{\boldsymbol{b}_{\mathrm{2}}},\mathbf{\boldsymbol{c}_{\mathrm{1}}}\right)-3P\left(\mathbf{\mathbf{\boldsymbol{a}}_{\mathrm{1}}\mathrm{,}\boldsymbol{b}_{\mathrm{2}}},\mathbf{\boldsymbol{c}_{\mathrm{1}}}\right)+P\left(\mathbf{\mathbf{\boldsymbol{a}}_{\mathrm{2}}\mathrm{,}\boldsymbol{b}_{\mathrm{2}}},\mathbf{\boldsymbol{c}_{\mathrm{1}}}\right)+3P\left(\mathbf{\boldsymbol{a}_{\mathrm{1}}},\mathbf{\boldsymbol{c}_{\mathrm{1}}}\right)-P\left(\mathbf{\boldsymbol{a}_{\mathrm{2}}},\mathbf{\boldsymbol{c}_{\mathrm{2}}}\right)+2P\left(\mathbf{\boldsymbol{b}_{\mathrm{1}}},\mathbf{\boldsymbol{c}_{\mathrm{2}}}\right)\nonumber \\
-4P\left(\mathbf{\mathbf{\boldsymbol{a}}_{\mathrm{1}}\mathrm{,}\boldsymbol{b}_{\mathrm{1}}},\mathbf{\boldsymbol{c}_{\mathrm{2}}}\right)-P\left(\mathbf{\mathbf{\boldsymbol{a}}_{\mathrm{2}}\mathrm{,}\boldsymbol{b}_{\mathrm{1}}},\mathbf{\boldsymbol{c}_{\mathrm{2}}}\right)-2P\left(\mathbf{\boldsymbol{b}_{\mathrm{2}}},\mathbf{\boldsymbol{c}_{\mathrm{2}}}\right)-P\left(\mathbf{\mathbf{\boldsymbol{a}}_{\mathrm{1}}\mathrm{,}\boldsymbol{b}_{\mathrm{2}}},\mathbf{\boldsymbol{c}_{\mathrm{2}}}\right)+3P\left(\mathbf{\mathbf{\boldsymbol{a}}_{\mathrm{2}}\mathrm{,}\boldsymbol{b}_{\mathrm{2}}},\mathbf{\boldsymbol{c}_{\mathrm{2}}}\right)\nonumber \\
\leq0,\;\label{eq:ch-ineq-3}
\end{eqnarray}
where for a tripartite correlated quantum system $\mathrm{ABC}$, $\mathbf{\boldsymbol{a}}_{1}$ and $\mathbf{\boldsymbol{a}}_{2}$ are two measurement directions of subsystem $\mathrm{A}$, $\mathbf{\boldsymbol{b}}_{1}$ and $\mathbf{\boldsymbol{b}}_{2}$ are the measurement directions of subsystem $\mathrm{B}$, $\mathbf{\boldsymbol{c}}_{1}$ and $\boldsymbol{\mathbf{c}}_{2}$ are the measurement directions of subsystem $\mathrm{C}$. Here $P\left(\cdots,\cdots,\cdots\right)$ denotes the probability of joint measurement of subsystems $\mathrm{A}$, $\mathrm{B}$ and $\mathrm{C}$, $P\left(\cdots,\cdots\right)$ is the joint probability when we measure two of the three subsystems simultaneously, and $P\left(\cdots\right)$ is the probability when we measure one of the three subsystems. Supposing the local hidden variable theory holds for the test quantum system, the maximum of Eq.~$\left(\ref{eq:ch-ineq-3}\right)$ cannot be larger than $0$.

\section{Entanglement and nonlocality inside a proton\label{sec:Nonlocality-and-entanglement}}

In this section we investigate the entanglement properties in the proton wave function. First, we concentrate on the entanglement between the valence quarks and the dynamical gluon inside a proton. Specifically, we consider two wave functions introduced in Section~\ref{subsec:BLFQ}: The first one involves solely the three valence quarks, denoted as $\left|qqq\right\rangle$, and the second one incorporates a dynamical gluon, denoted as $\left|qqq\right\rangle +\left|qqqg\right\rangle$. For each wave function, we present the results of the entanglement entropy for the spin and longitudinal momentum of the individual partons and the remainder of the proton, where we perform partial trace over all the other degrees in the proton wave function than the spin and longitudinal momentum respectively. In addition, we evaluate the maximum violation of the Bell-CH inequality Eq.~$\left(\ref{eq:ch-ineq-3}\right)$ using the spin density matrix of the valence quarks $\rho_{qqq}$. In the scenario involving only the $\left|qqq\right\rangle$ state, we observe a modest violation of Eq.~$\left(\ref{eq:ch-ineq-3}\right)$. However, in the mixed state $\rho_{qqq}$ reduced from $\left(\left|qqq\right\rangle +\left|qqqg\right\rangle\right)\left(\left\langle qqq\right| +\left\langle qqqg\right|\right)$, the nonlocal correlations among the three quarks do not exhibit any detectable inequality violations. This difference reflects the underlying pattern of the information exchange between quarks and gluons and that of the quantum entanglement within the proton structure.

\subsection{The state $\left|qqq\right\rangle$\label{subsec:qqq-flavor}}

Here the wave function $\left|qqq\right\rangle$ is computed in the framework of BLFQ. There are in total $8$ kinds of spin configurations and $136$ kinds of longitudinal momentum configurations. Here, we use the simulator backend \textit{Qton} of the quantum computing cloud platform \textit{Quafu} developed from Beijing Academy of Quantum Information Sciences to perform partial trace in the density matrix over all the other degrees of freedom than the spin and longitudinal momentum which we concentrate on and calculate entanglement entropy~\cite{Xu_2024}. We need at least $\mathrm{\log_{2}}{8}=3$ qubits to encode the spin state and $\mathrm{\log_{2}}{256}=8$ qubits for the longitudinal momentum state. For spin states, we use $\left\{ \left|0\right\rangle,\left|1\right\rangle \right\}$ to denote $\left\{ \left|\downarrow\right\rangle_{q},\left|\uparrow\right\rangle _{q}\right\}$. In the configuration of longitudinal momentum, the value range of the three constituents must be from $0.5$ to $K_{max}-1=15.5$, and the sum of the three constituents is $16.5$. We allocate $4$ qubits for each constituent to represent binary numbers from $0.5$ to $15.5$, resulting in a total of $3\times4=12$ qubits required for the computation. On the whole, the formation of the density matrix of $\left|qqq\right\rangle$ is
\begin{equation}
\rho_{qqq}=\left|q_{1}q_{2}q_{3}\right\rangle \left\langle q_{1}q_{2}q_{3}\right|,\;\label{eq:qqq-dm}
\end{equation}
where the flavor of $q_{1}$ and $q_{3}$ is $u$ quarks, and $q_{2}$ is a $d$ quark.
\noindent \begin{center}
\begin{figure}
\noindent \begin{centering}
\includegraphics{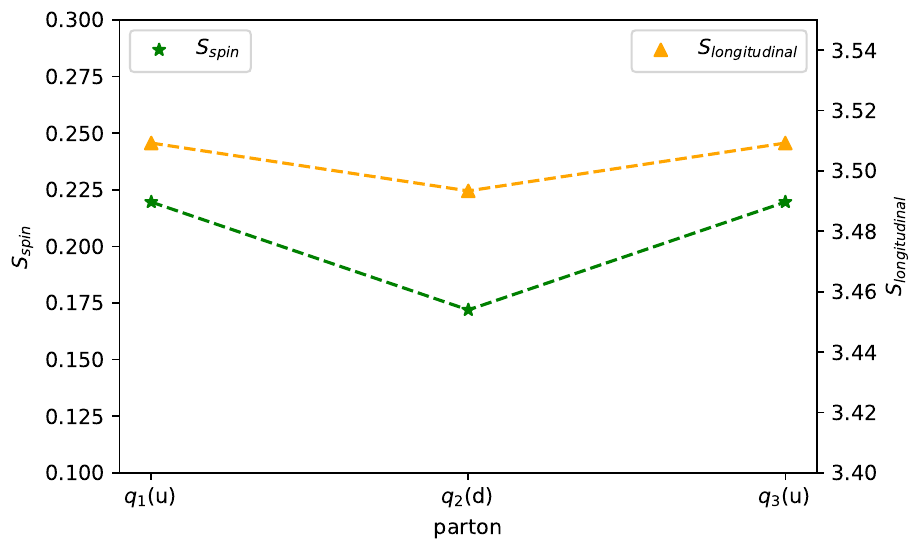}
\par\end{centering}
\caption{\label{fig:qqq-ent}\footnotesize The entanglement entropy of spin and longitudinal momentum between each valence quark and the remaining part of the BLFQ proton in the wave function $\left|qqq\right\rangle$. The green dots denote the entropy of the spin states $S_{\mathrm{spin}}$ for $q_{1}$, $q_{2}$ and $q_{3}$: $\left\{ 0.219,0.172,0.219\right\}$. The orange dots denote the entropy of the longitudinal momentum $S_{\mathrm{longitudinal}}$ for $q_{1}$, $q_{2}$ and $q_{3}$: $\left\{ 3.510,3.493,3.510\right\} $.}
\end{figure}
\par\end{center}

We present the entanglement entropy results for both the spin and longitudinal momentum states in Fig.~$\ref{fig:qqq-ent}$, , which shows that the wave functions of the two $u$ quarks are symmetric, resulting in the equal entanglement for the two $u$ quarks with the remaining $ud$ pairs, while the entanglement for between the $d$ quark and the remaining $uu$ pair is comparatively lower. As the maximal value of entanglement entropy for one spin-$\frac{1}{2}$ system with its complementary parts is $1$, the spin entanglement among valence quarks is generally weak. It implies that the direct information exchange between the spin of the valence quarks is weak. The entanglement of longitudinal momentum is relatively large, since each constituent is represented by $4$ qubits, and the maximal entanglement entropy is $\log_{2}2^{4}=4$. The sizable entanglement in the longitudinal momentum may partly be explained by the classical correlation between the longitudinal momentum of the three quarks: the proton wave function demands its total longitudinal momentum to be $16.5$, which brings a classical correlation between the three constituents. 

Next, we take the spin state of $\left|qqq\right\rangle$ to test the violation of Bell-CH inequality in Eq.~$\left(\ref{eq:ch-ineq-3}\right)$, and the numerical maximal violation is $0.0905553$. This violation suggests that in addition to the entanglement, the Bell nonlocal correlation exists among the three valence quarks inside the proton.

\subsection{The state $\left|qqq\right\rangle +\left|qqqg\right\rangle$\label{subsec:qqqg-flavor}}

By including one dynamic gluon in the BLFQ wave function, we can study quantum correlations both between valence quarks and between quarks and gluons. The wave function of $\left|qqq\right\rangle +\left|qqqg\right\rangle$ has in total $24$ spin configurations and $816$ longitudinal momentum configurations. To characterize the spin of the gluon, we need $2$ more qubits compared with $\left|qqq\right\rangle$. Here we use $\left\{ \left|00\right\rangle ,\left|01\right\rangle ,\left|10\right\rangle \right\}$ to denote the gluon spin configuration $\left\{ \left|\mathrm{\Omega}\right\rangle_{g},\left|\downarrow\right\rangle_{g}, \left|\uparrow\right\rangle_{g} \right\}$, where $\left|\mathrm{\Omega}\right\rangle_{g}$ denotes the state with no gluon. Consequently, $\mathrm{\log_{2}}{8}+2=5$ qubits suffice to characterize the spin state. The longitudinal momentum state of $\left|qqq\right\rangle +\left|qqqg\right\rangle$ has four constituents with a sum of $16.5$ (the minimal longitudinal momentum of the gluon is $1$). Then the total number of qubits we need to describe the longitudinal momentum state is $4\times4=16$, and the fourth constituent is set to be zero for the $\left|qqq\right\rangle$ Fock sector. Schematically, we can expand the density matrix of $\left(\left|qqq\right\rangle +\left|qqqg\right\rangle\right)\left(\left\langle qqq\right| +\left\langle qqqg\right|\right)$ as
\begin{equation}
\rho_{qqqg}=\left[\begin{array}{cc}
\left|q_{1}q_{2}q_{3}\right\rangle \left\langle q_{1}q_{2}q_{3}\right| & \left|q_{1}q_{2}q_{3}\right\rangle \left\langle q_{1}q_{2}q_{3}g\right|\\
\left|q_{1}q_{2}q_{3}g\right\rangle \left\langle q_{1}q_{2}q_{3}\right| & \left|q_{1}q_{2}q_{3}g\right\rangle \left\langle q_{1}q_{2}q_{3}g\right|
\end{array}\right],\;\label{eq:qqqg-dm}
\end{equation}
where the flavors of $q_{1}$, $q_{2}$ and $q_{3}$ are $u$, $d$ and $u$ quarks, respectively. As the mediator of the strong interaction, the gluon $g$ is expected to exchange information among the three quarks.
\noindent \begin{center}
\begin{figure}
\noindent \begin{centering}
\includegraphics{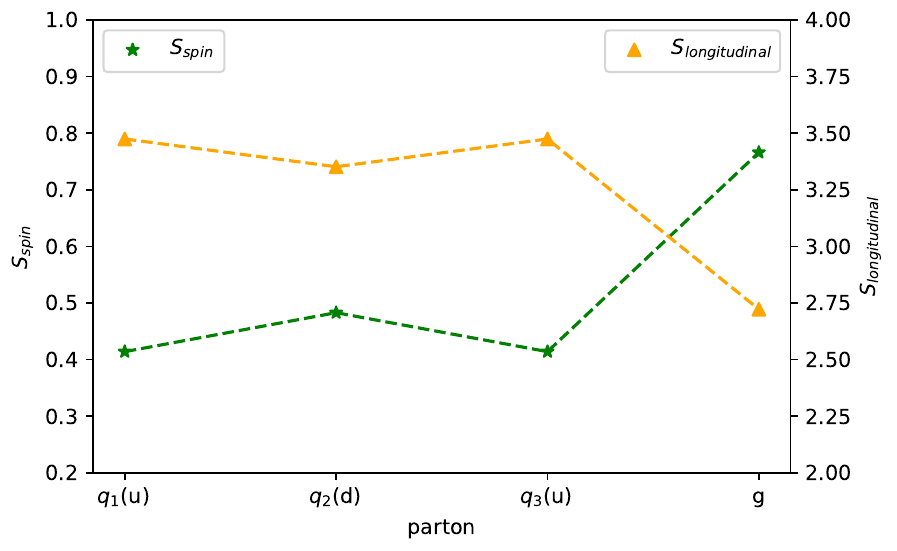}\caption{\label{fig:qqqg-ent}\footnotesize The entanglement entropy of spin and longitudinal momentum between each valence quark and the remaining part of the BLFQ proton wave function $\left|qqq\right\rangle +\left|qqqg\right\rangle$. The green dots denote the entropy of the spin states $S_{\mathrm{spin}}$ for $q_{1}$, $q_{2}$, $q_{3}$, and $g$: $\left\{ 0.414,0.483,0.414,0.766\right\}$.
The orange dots denote the entropy of the longitudinal momentum $S_{\mathrm{longitudinal}}$ for $q_{1}$, $q_{2}$, $q_{3}$, and $g$: $\left\{ 3.474,3.352,3.474,2.722\right\}$.}
\par\end{centering}
\end{figure}
\par\end{center}

By taking partial trace of $\rho_{qqqg}$ individually respect to $q_{1}$, $q_{2}$, $q_{3}$ and $g$ via Eq.~$\left(\ref{eq:trace}\right)$, and computing entanglement entropy via Eq.~$\left(\ref{eq:Shannon-entropy}\right)$, we present the results in Fig.~$\ref{fig:qqqg-ent}$. The maximal entanglement entropy is $1$ and $4$ for spin and longitudinal momentum, respectively. From Fig.~$\ref{fig:qqqg-ent}$, we find the spin entanglement of the gluon with remaining part of the proton is relatively larger than that of a quark in the state $\left|qqq\right\rangle$. Besides, the longitudinal momentum entanglement seems still large among quarks but relatively weak between the gluon and quarks. This is because when there is no gluon, the fourth constituent should be $0$, which is excluded from the classical constraint of sum $16.5$.

Since the quark spin is a qubit whereas the gluon spin is a qutrit, we have not derived an appropriate Bell-CH inequality for the whole system of $\left|qqq\right\rangle +\left|qqqg\right\rangle$. Therefore, here we restrict to the exploration of the Bell nonlocal correlations between spins of the three valence quarks $\rho_{qqq}$ by tracing over the gluon $g$ in $\rho_{qqqg}$. In this case we observe no violation from Eq.~$\left(\ref{eq:ch-ineq-3}\right)$, and the maximum value calculated was $-0.209525$. One possible explanation is that the Bell-CH inequalities are relatively not sensitive to mixed states, and another possibility is that a significant part of the quantum correlation is carried through the gluon according to Fig.~$\ref{fig:qqqg-ent}$ and with the gluon degrees of freedom being traced over, the remaining quantum correlation between the quarks becomes weak.


Overall, comparing all the results from $\left|qqq\right\rangle$ and $\left|qqq\right\rangle+\left|qqqg\right\rangle$, we find that the involvement of one dynamical gluon significantly enhances the spin entanglement within the proton. Specifically, the gluon acting as a mediator among the three valence quarks, facilitates the exchange of quantum information in the proton. The quantum information protocol of quark-gluon is worth further study to understand the role of gluons inside the proton.

\section{Spin entanglement for different longitudinal momentum\label{sec:Spin-entanglement}}

Having studied the quantum correlations among different partons inside the BLFQ proton wave function, we turn to another perspective in this section. In hadron scattering experiments, detecting the spin or longitudinal momentum of a specific single quark or gluon is difficult. Instead, one often measues the helicity parton distribution functions of the proton, which encode the information on the spin carried by the partons with given longitudinal momentum. Therefore, in this section we study the spin entanglement entropy in terms of the longitudinal momentum of partons in this section. Likewise, we consider two scenarios: $\left|qqq\right\rangle $ and $\left|qqq\right\rangle +\left|qqqg\right\rangle$.

\subsection{The state $\left|qqq\right\rangle $\label{subsec:qqq-x}}

In the wave function with only the $\left|qqq\right\rangle$ sector, both the longitudinal momentum and spin parts have three constituents. In the previous studies of parton entanglement entropy in DIS~\cite{Kharzeev:2017qzs, Tu:2019ouv, Gotsman:2020bjc, Kharzeev:2021yyf, Zhang:2021hra}, the authors considered two situations according to the partons' longitudinal momentum fraction $x$: when $x$ is small, the dominant contribution is from gluons; when $x$ is large, the contribution of sea quarks should be taken into consideration. Based on the similar idea, we divide partons inside the BLFQ proton into small, intermediate and large $x$ regions according to the values of their longitudinal momenta:
\begin{equation}
\left|x_{1},x_{2},x_{3}\right\rangle _{\mathrm{longitudinal}}\otimes\left|s_{1},s_{2},s_{3}\right\rangle _{\mathrm{spin}}\Rightarrow\left|x_{\mathrm{min}},x_{\mathrm{mid}},x_{\mathrm{max}}\right\rangle _{\mathrm{longitudinal}}\otimes\left|s_{\mathrm{min}},s_{\mathrm{mid}},s_{\mathrm{max}}\right\rangle _{\mathrm{spin}},\;\label{eq:rearrange}
\end{equation}
where we have $x_{1}+x_{2}+x_{3}=1$, along with $x_{\mathrm{min}}=\mathrm{min}\left\{ x_{1},x_{2},x_{3}\right\}$, $x_{\mathrm{max}}=\mathrm{max}\left\{ x_{1},x_{2},x_{3}\right\}$ and $x_{\mathrm{mid}}$ denotes the middle value of $x_{1,2,3}$. When we exchange the longitudinal momentum constituents according to $x$, we exchange the spin constituents correspondingly. Thus, we have $\left\{ s_{1},s_{2},s_{3}\right\} \rightarrow\left\{ s_{\mathrm{min}},s_{\mathrm{mid}},s_{\mathrm{max}}\right\}$. This rearrangement allows us to compare the spin entanglement of the small $x$ region and the large $x$ region.
After the rearrangement, we encode the three spin constituents using $8$ qubits similar to Sec.~\ref{subsec:qqq-flavor} and obtain
the reduced density matrices
\begin{equation}
\rho_{\mathrm{min}}=\mathrm{Tr}_{\mathrm{mid}\,\mathrm{max}}\rho_{qqq}^{r},\:\rho_{\mathrm{max}}=\mathrm{Tr}_{\mathrm{min}\,\mathrm{mid}}\rho_{qqq}^{r},\;\label{eq:reduced min-max}
\end{equation}
where $\rho_{qqq}^{r}=\left|q_{\mathrm{min}}q_{\mathrm{mid}}q_{\mathrm{max}}\right\rangle \left\langle q_{\mathrm{min}}q_{\mathrm{mid}}q_{\mathrm{max}}\right|$. Since we aim to compute spin entanglement entropies of different $x$ regions via Eq.~$\left(\ref{eq:Shannon-entropy}\right)$, we define ``small $x$ region'' as $x_{\mathrm{min}}\leq 5.5/15.5$ and ``large $x$ region'' as $x_{\mathrm{max}}\geq 5.5/15.5$. As an exploratory study of the correlation between small-(large-)$x$ partons and the remaining part of the proton, we single out the configurations with $x_\text{min}\le 5.5/15.5$ ($x_\text{min}\ge 5.5/15.5$) from the BLFQ wave function to form a new wave function, the norm of which is renormalized to $1$. Next, we study the entanglement between partons in different $x$ regions based on these new wave functions. First, we take the trace of the new wave functions according to Eq.~$\left(\ref{eq:reduced min-max}\right)$. The spin entanglement entropy for small or large $x$ regions has been depicted in Fig.~$\ref{fig:qqq-max-min}$, where we find that for $x_{\mathrm{min}}\leq X_{\mathrm{min}}$, the entanglement increases as $X_{\mathrm{min}}$ grows; whereas, for $x_{\mathrm{max}}\geq X_{\mathrm{max}}$, the entanglement generally decreases as $X_{\mathrm{max}}$ grows. This can be explained by the fact that the entanglement entropy in Eq.~$\left(\ref{eq:reduced min-max}\right)$ depends on both the number and amplitudes of the configurations involved. With larger $X_{\mathrm{min}}$ or smaller $X_{\mathrm{max}}$, the new set of wave functions include more configurations from the original proton wave function, which make the probability of entanglement increase.

\noindent \begin{center}
\begin{figure}
\noindent \begin{centering}
\includegraphics{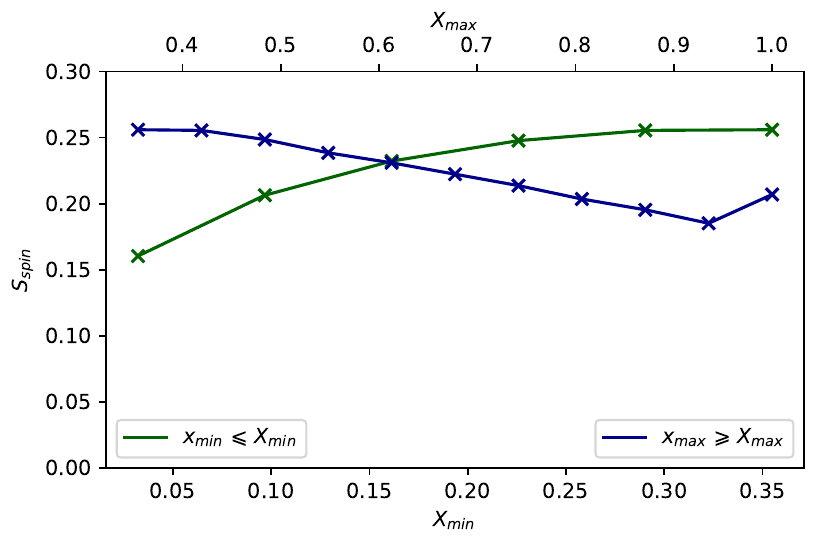}\caption{\label{fig:qqq-max-min}\footnotesize The spin entanglement entropy between the parton $x_{\mathrm{min}}\protect\leq5.5/15.5$, $x_{\mathrm{max}}\protect\geq5.5/15.5$ and the remaining partons. The entanglement of a small $x$ region is the dark green dots and line, and that of a large $x$ region is the dark blue dots and line. }
\par\end{centering}
\end{figure}
\par\end{center}

After analyzing the entanglement entropy of small or large $x$ regions, we are also interested in more details of the spin entanglement entropy of partons with a particular longitudinal momentum $x_{f}$. For the convenience of evaluating entanglement entropy, we exchange the parton with this particular longitudinal momentum value to a fixed place. The permutation becomes as follows
\begin{equation}
\left|x_{1},x_{2},x_{3}\right\rangle _{\mathrm{longitudinal}}\otimes\left|s_{1},s_{2},s_{3}\right\rangle _{\mathrm{spin}}\Rightarrow\left|x_{f},x_{2\leftrightarrow1},x_{3\leftrightarrow1}\right\rangle _{\mathrm{longitudinal}}\otimes\left|s_{f},s_{2\leftrightarrow1},s_{3\leftrightarrow1}\right\rangle _{\mathrm{spin}},\;\label{eq:exchange-fixed}
\end{equation}
where the fixed place is chosen to be the first place in the ket, and $x_{f}$ has the value from $0.5/15.5$ to $15.5/15.5$. Similar to the above procedure, we first single out all configurations from the original proton wave function including the value $x_{f}$, and form a new wave function by renormalizing the new wave function to $1$. Then, we exchange $x_{f}$ in the configurations with $x_{1}$ (no exchange needed when $x_{f}=x_{1}$), and permute all the other quantum numbers including the spin $s_{f}$ and $s_{1}$ correspondingly, which is shown in Eq.~$\left(\ref{eq:exchange-fixed}\right)$. For example, if $x_{f}=x_{3}$, we exchange the place of $x_{1}$ and $x_{3}$ as $\left|x_{1},x_{2},x_{3}\right\rangle \otimes\left|s_{1},s_{2},s_{3}\right\rangle \left|x_{3},x_{2},x_{1}\right\rangle \otimes\left|s_{3},s_{2},s_{1}\right\rangle$. After the exchanges, we perform the trace over the quantum numbers of the second and third constituent and obtain a reduced density matrix $\rho_{f}$
\begin{equation}
\rho_{f}=\mathrm{Tr}_{\mathrm{other\,than}\,f}\rho_{qqq}^{f}\;\label{eq:fixed-rho}
\end{equation}
where $\rho_{qqq}^{f}=\left|q_{f}q_{2\leftrightarrow1}q_{3\leftrightarrow1}\right\rangle \left\langle q_{f}q_{2\leftrightarrow1}q_{3\leftrightarrow1}\right|$ denotes the reduced spin density matrix of the longitudinal momentum with the fixed value $x_{f}$. Based on $\rho_f$, we calculate the entanglement entropy of the parton with longitudinal momentum $x_f$ with the remaining part of the proton and present the results in Fig.~$\ref{fig:qqq-fixed}$. The trend of $S_{\mathrm{spin}}$ as a function $x_{f}$ is shown in Fig.~$\ref{fig:qqq-fixed}$, which is determined by the number and amplitudes of the configurations containing a parton with the longitudinal momentum fraction $x_f$. $S_\text{spin}$ increases $x_f$, which seems consistent with the results in Fig.~$\ref{fig:qqq-max-min}$. This implies that $S_\text{spin}$ for the small- (large-) $x$ region is strongly influence by the partons with largest (smallest) $x_f$ in that region.
\noindent \begin{center}
\begin{figure}
\noindent \begin{centering}
\includegraphics{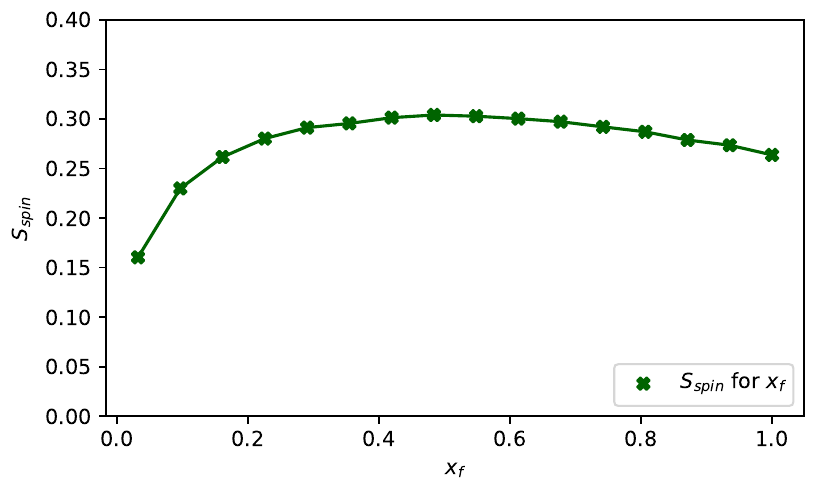}\caption{\label{fig:qqq-fixed}\footnotesize The entanglement entropy between the parton $x_{f}=0.5/15.5$ to $15.5/15.5$ and rest partons. Here the dark green dots and line show the spin entanglement entropy in terms of $x_{f}$.}
\par\end{centering}
\end{figure}
\par\end{center}

\subsection{The state $\left|qqq\right\rangle +\left|qqqg\right\rangle$\label{subsec:qqqg-x}}

By including a dynamical gluon into the BLFQ proton wave function, we can study the entanglement between quarks and gluons. As the four constituents of the longitudinal momentum represent different types of partons, it is difficult to define the rearrangement in terms of small or large $x$ like Eq.~$\left(\ref{eq:rearrange}\right)$. However, it offers us a chance to study the entanglement contribution of the gluon to the whole proton with a certain longitudinal momentum $x_{f}$. First, we consider the entanglement of one quark with the remaining part. We exchange the selected quark with certain $x_{f}$ to the first constituent as Eq.~$\left(\ref{eq:exchange-fixed}\right)$. This permutation has the form of
\begin{equation}
\left|x_{1},x_{2},x_{3},x_{g}\right\rangle _{\mathrm{longitudinal}}\otimes\left|s_{1},s_{2},s_{3},s_{g}\right\rangle _{\mathrm{spin}}\Rightarrow\left|x_{f},x_{2\leftrightarrow1},x_{3\leftrightarrow1},x_{g}\right\rangle _{\mathrm{longitudinal}}\otimes\left|s_{f},s_{2\leftrightarrow1},s_{3\leftrightarrow1},s_{g}\right\rangle _{\mathrm{spin}},\;\label{eq:qqqg-q-fixed}
\end{equation}
where all definitions are the same as Eq.~$\left(\ref{eq:exchange-fixed}\right)$, $x_{g}$ denotes the longitudinal momentum of dynamical gluon with $x_{1}+x_{2}+x_{3}+x_{g}=1$, and $s_{g}$ is the corresponding spin of the gluon. Both $x_{g}$ and $s_{g}$ equals $0$ for Fock sectors $\left|qqq\right\rangle$, where there is no gluon. Next, we consider the entanglement of one dynamical gluon with the remaining quark part. Since there is only one gluon in the original proton wave function, it is easier to arrange the fourth constituent $x_{g}$ from $1/15$ to $15/15$ as
\begin{equation}
\left|x_{1},x_{2},x_{3},x_{g}\right\rangle _{\mathrm{longitudinal}}\otimes\left|s_{1},s_{2},s_{3},s_{g}\right\rangle _{\mathrm{spin}}\Rightarrow\left|x_{1},x_{2},x_{3},x_{f}\right\rangle _{\mathrm{longitudinal}}\otimes\left|s_{1},s_{2},s_{3},s_{f}\right\rangle _{\mathrm{spin}}.\;\label{eq:qqqg-g-fixed}
\end{equation}
Given all these conditions, we present the comparison results in Fig.~$\ref{fig:qqqg-fixed}$. In order to better understand the role played by the gluon in terms of entanglement entropy, we perform the same entanglement entropy calculation for a dressed quark using the wave function solved in $\left|q\right\rangle +\left|qg\right\rangle$ for comparison. The parameters used in the dressed quark calculation are the same as those in the proton, with the only exceptions being the truncation parameters $\{N_\text{max},K\}$=\{7, 15.5\} instead of \{9, 16.5\}. The reason is that the proton has three quarks and each quark will at least occupy one of transverse quanta in the $N_\text{max}$ counting and $0.5$ unit in the longitudinal momentum, so the dressed quark with the truncation parameters $\{N_\text{max},K\}$=\{7, 15.5\} can best simulate a quark ``embeded'' in the proton, which is shown in the inset of Fig.~$\ref{fig:qqqg-fixed}$.
\noindent \begin{center}
\begin{figure}
\noindent \begin{centering}
\includegraphics{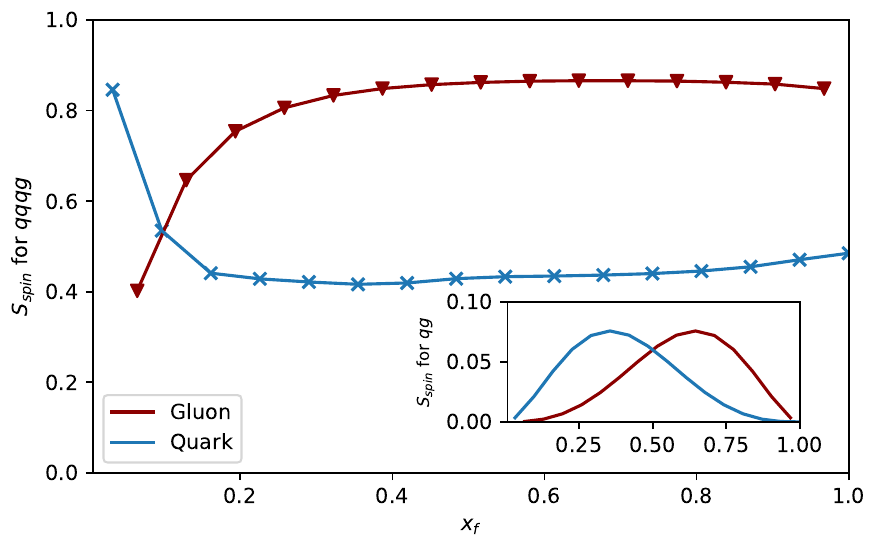}
\par\end{centering}
\caption{\label{fig:qqqg-fixed} \footnotesize Comparison of the spin entanglement entropy of quark and gluon with fixed longitudinal momentum $x_{f}$ inside the proton wave function $\left|qqq\right\rangle +\left|qqqg\right\rangle$. Meanwhile, the inset shows the entanglement entropy of quarks and gluons inside the dressed quark $\left|q\right\rangle +\left|qg\right\rangle$. The blue dots and line show entanglement entropy between the selected quark and the remaining part, and the dark red dots and line show entanglement entropy between the selected gluon and the remaining quarks.}
\end{figure}
\par\end{center}

We can extract several important features from Fig.~$\ref{fig:qqqg-fixed}$. Firstly, the entanglement entropy $S_{\mathrm{spin}}$ exhibits an approximate symmetry concerning $x_{f}$ in both proton and dressed quark systems. This means, the value of $S_{\mathrm{spin}}$ at $x_{f}=x_{0}$ for the selected quark is roughly equal to $S_{\mathrm{spin}}$ when $x_{f}=1-x_{0}$ for the dynamical gluon in the wave function $\left|qqq\right\rangle +\left|qqqg\right\rangle$, and they are precisely equal in the wave function of dressed quark $\left|q\right\rangle +\left|qg\right\rangle$.
Secondly, we find $S_{\mathrm{spin}}$ of quarks and the dynamical gluon in $\left|qqq\right\rangle +\left|qqqg\right\rangle$ is much larger than $S_{\mathrm{spin}}$ in $\left|q\right\rangle +\left|qg\right\rangle$. Notice that in the dressed quark, the gluon only acts on self-energy correction, but the gluons inside a proton have two roles: self-energy correction and quark-quark exchange interaction. In this model, since $S_{\mathrm{spin}}$ for the dressed quark is small, we conjecture that gluons acting as self-energy correction do not contribute a lot of entanglement. That is, most entanglement of the dynamical gluon in the proton arises from exchange interactions among quarks. In other words, analyzing entanglement between quarks and gluons may bring us a novel method to distinguish the role of gluons inside a proton or other hadrons.

\section{Conclusions and outlook\label{sec:Conclusions-and-outlook}}

In this work, we investigate the entanglement properties of quarks and gluons in two light-front wave functions for the proton from BLFQ, which are respectively evaluated in the $\left|qqq\right\rangle$ and $\left|qqq\right\rangle +\left|qqqg\right\rangle$ Fock space. We analyze the entanglement entropy in the spin and longitudinal degrees of freedom of partons and study the quantum nonlocality through violation of the Bell-CH inequality of tripartite spin states $\rho_{qqq}$ for both wave functions. We find that the spin entanglement for the wave function in the $\left|qqq\right\rangle +\left|qqqg\right\rangle$ Fock space is larger than that in $\left|qqq\right\rangle$, which indicates the presence of one dynamical gluon increases the quantum correlation between the quark and remaining part of the proton. In addition, we study the spin entanglement entropy between a parton with a fixed longitudinal momentum fraction and the remaining part of the proton. We compare the results with those for a dressed quark state $\left|q\right\rangle +\left|qg\right\rangle$ and find that in the dressed quark the entanglement entropy between the gluon and the remaining part of the system is much smaller than that in the proton.  These results suggest that the exchange gluon between the quarks may generate larger amount of entanglement entropy compared to that acting in self-energy corrections. Finally, we point out that the entanglement entropy of quarks and gluons may experimentally accessible through the measurement of their helicity distribution functions.

As the next step, we plan to extend this work by considering the proton wave functions including higher Fock sectors, such as those with sea quarks and more than one dynamical gluon~\cite{Xu:2024sjt}. From these wave functions we can further study the entanglement between valence quarks, gluons and sea quarks, which will hopefully provide a more complete picture of quantum correlation in the proton. Since the wave functions with higher Fock sectors correspond to higher energy scales, these results will allow for more direct comparison with the experimental measurements of entanglement through parton distribution functions. Besides the proton, we can also apply this study to other hadrons to obtain a comprehensive understanding of hadron structure in terms of quantum entanglement. 

\begin{acknowledgments}
We thank Yun-Heng Ma for \emph{Qton} software supports. This work is supported by the National Natural Science Foundation of China (NSFC) with Grant No. 12305010.
\end{acknowledgments}

\bibliography{refs-ent-blfq}
\end{document}